\documentclass{article}
\usepackage[affil-it]{authblk} 
\usepackage{textcomp, amssymb}

\usepackage[utf8]{inputenc}
\usepackage[dvipsnames]{xcolor}
\usepackage{amsmath}
\usepackage{graphicx}
\usepackage{lmodern}
\usepackage{amssymb}
\usepackage{subfig}
\usepackage{upgreek}

\usepackage[square,comma,sort&compress,numbers]{natbib}
\providecommand{\keywords}[1]
{
  \small	
  \textbf{\textit{Keywords---}} #1
} 
\usepackage{soul}
\usepackage{natbib}
\usepackage{xcolor}

\begin{document}
\title{Plasmonic tweezers based on connected nanoring apertures}
\author{Theodoros D. Bouloumis}
\author{Domna G. Kotsifaki}
\author[2] {Xue Han}
\author[1,*]{Síle {Nic Chormaic}}
\author[1]{Viet Giang Truong}

\affil[1]{Light-Matter Interactions for Quantum Technologies Unit, Okinawa Institute of Science and Technology Graduate University, Onna-San, Okinawa, Japan }
\affil[2]{School of Optoelectronic Engineering and Instrumentation Science, Dalian University of Technology, Dalian, Liaoning, China }
\affil[*]{sile.nicchormaic@oist.jp}
\date{} 

\maketitle

\begin{abstract}
The  manipulation of microparticles using optical forces has led to many applications in the life and physical sciences. To extend optical trapping towards the nano-regime, in this work we demonstrate  trapping of single nanoparticles in arrays of plasmonic coaxial nano-apertures with various inner disk configurations and theoretically estimate the associated forces. A high normalised experimental trap stiffness of 3.50 fN/nm/mW for 20~nm polystyrene particles is observed for an optimum design of 149~nm for the nanodisk diameter at a trapping wavelength of 980~nm. Theoretical simulations are used to interpret the enhancement of the observed trap stiffness. A quick particle trapping time of less than 8~sec is obtained at a concentration of 14~$\times$~10$^{11}$ particles/ml with low incident laser intensity of 0.59~mW/$\mu$m$^{2}$. This good trapping performance with fast delivery of nanoparticles to multiple trapping sites emerges from a combination of the enhanced electromagnetic near-field and spatial temperature increase. This work has applications in nanoparticle delivery and trapping with high accuracy, and bridges the gap between optical manipulation and nanofluidics.

\end{abstract}

\newpage
\section{Introduction} 

Optical tweezers are a powerful tool to trap and manipulate particles whose size ranges from a few hundred nanometres to tens of micrometres~\cite{R15}. Despite numerous applications, optical trapping of particles with sizes smaller than the wavelength of light remains a considerable challenge due to the diffraction limit, which hinders the trapping process. Pushing optical trapping into the subwavelength regime is achieved with plasmonic optical tweezers (POTs) \cite{R3,R1,Domna,R2, Marago}, ushering in a new era of experiments at the single nanoparticle level. POTs rely on subwavelength confinement of the trapping laser beam into highly intense hotspots. Complex metallic nanostructures are used to create such plasmonic hotspots, which strongly enhance the amplitude of the electric field of the incident trapping laser. This increased electric field compensates for the low polarisability of nanosized particles, resulting in narrower and deeper trapping potential wells compared to conventional optical tweezers.

 A critical issue for  trapping individual nanoparticles in an aqueous environment is how to manipulate them  towards a given localised near-field hotspot, where they can be trapped by optical gradient forces~\cite{R20,R31}. To address this problem, there are different approaches that primarily enable transport and trapping with plasmonic nano-antennas. Among these techniques, periodic arrays of closely spaced nano-antennas, e.g., arrays of metallic nanodisks~\cite{R14}, bowtie  nanocavities~\cite{R8}, gold nanopyramidal-dimers~\cite{R21}, etc., have gained considerable interest. In all cases, collective heating, due to nearby hotspots, produces strong fluidic thermal convection and thermophoresis, which exert drag forces on the suspended particles and impact their motion~\cite{R8, R20}. This intrinsic thermoplasmonic convection is exploited for fluid transport~\cite{R8, R21, R22}; hence, particles can mobilise into the area illuminated by the incident laser with  relatively fast fluid motion (1~$\mu$m/s)~\cite{R32}. Nevertheless, this method suffers from the issue of ensembles of particles or agglomeration over the surface of the metallic nanostructured array~\cite{R8,R21, R22}, leading to  difficulty in addressing single nanoparticle trapping at a given hotspot.
 
A different approach for trapping single nano-objects involves the illumination of a single plasmonic nanostructured element, i.e., a single or a double-nanohole on a metallic thin film~\cite{{Pang, Xu}}. In this approach, collective heating is absent and the intrinsic thermoplasmonic convection is very weak, i.e., the convection fluid velocity is less than 10 nm/s~\cite{Xu,R27,R28}. Therefore, the trapping process is mainly due to the optical gradient force~\cite{R26, R29}. Experimentally, the trapping of individual nano-objects (smaller than 10 nm in diameter) has been demonstrated at relatively low laser intensities of less than 1 mW/$\mu$m$^{2}$~\cite{Xu,R5,Pang,Pang2,R26}. In spite intense interest in the trapping of nanoparticles with low light trapping powers, the movement of the suspended particles relies on Brownian motion and is diffusion-limited~\cite{R26,R28}. Consequently, only particles in close proximity to the plasmonic nanostructure can be effectively trapped, placing a limit on working with low concentration and high speed particle trapping applications~\cite{R20,R31}.  Another approach to transport particles to a specific plasmonic hotspot is by integrating an AC electric field with plasmonic nano-antenna resonators~\cite{R20}. However, the electro-thermo-plasmonic  flow associated with the small temperature gradient opto-thermally generated at a single plasmonic nano-antenna is too weak for effective particle delivery.

Recently, we proposed and demonstrated arrays of nanoring apertures for multiple particle trapping for long trapping periods using low incident laser powers~\cite{R9,Han2,Han}. 
The trapping site density was increased by introducing 50 nm connecting gaps between neighbouring nanoring apertures. Using such a  device, we trapped and transported polystyrene (PS) particles of 0.5~$\mu$m and 1~$\mu$m diameter across the device's surface with a trapping laser intensity of 1.5~mW/$\mu$m$^{2}$~\cite{Han2}. To trap single nanoparticles (30~nm, dielectric), a modified double nanohole array was illuminated by light with an intensity of 0.51 mW/$\mu$m$^{2}$ for on-resonance conditions~\cite{Han}. Most recently, we fabricated arrays of asymmetric, Fano  resonance, split nano-apertures for  trapping single 20 nm PS particles on a 50 nm gold thin film. A very large normalised trap stiffness of 8.65~fN/nm/mW was achieved ~\cite{R4}. An important feature in using these double nanohole/Fano resonance POT aperture arrays is that one can trap individual nanoparticles at low incident powers. Not only does our POT system largely improve the trap stiffness, it also reduces diffusion-limited trapping. The average diffusion time to mobilise particles towards the illuminated trapping area reduces from 15 - 20 minutes for a typical single nano-element POT ~\cite{Xu} to approximately 60 s for the POT aperture array ~\cite{R4,Han} at a similar incident laser intensity ($\sim$1 mW/$\mu$m$^{2}$) and particle concentration of 14~$\times$~10$^{11}$ particles/ml~\cite{Kotnala,Xu} in water solution.

In this paper, we theoretically calculate the optical forces and trap stiffnesses for a plasmonic optical tweezers  
and experimentally demonstrate an efficient optical tweezers for 20 nm PS particle trapping. We use a similar POT design as before~\cite{Han2}. By introducing coaxial inner disks, the trap stiffness for nanoparticles increases by a factor of about three  compared to that for our previous nanohole array ~\cite{Han}. The average time until the first observed trapping event is reduced from 20~s (nanohole structure) to 8~s (nanohole plus disk structure) for a low laser intensity of 0.59~mW/$\mu$m$^{2}$ and a concentration of 14~$\times$~10$^{11}$ particles/ml.
With quick delivery of particles to the aperture array surface, while maintaining control of single nanoparticles at specific trapping nanoslot antennas, this POT has the potential to trap and deliver nanoparticles or analytes to various locations on a chip environment with high accuracy. This opens the way for the development of a non-destructive and highly sensitive tool for single molecule detection and studies on specific trapping positions.

\section{Results and Discussion} 

\subsection{Absorption resonance and optical force calculation}

\begin{figure}
\centering
   \includegraphics[trim={0.5cm 0cm 0.2cm 0.5cm},clip, width=12cm, height= 7cm]{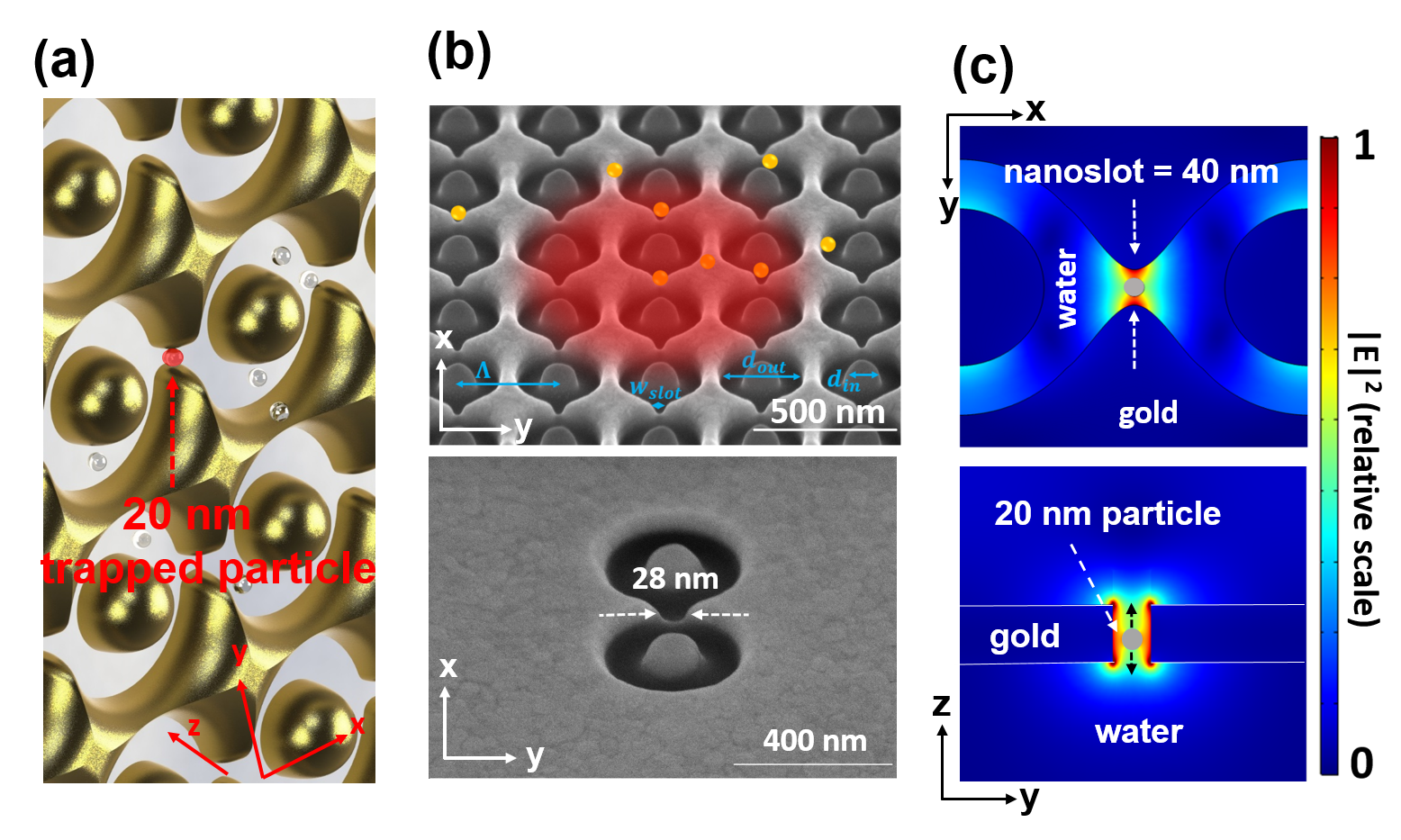}
   \caption{(a) Schematic of an array of nanoring nano-apertures.(b) Scanning electron microscope (SEM) images of an array of fabricated nano-apertures in which the sample is titled by 52$^{\circ}$ from the surface normal. The outer ring diameter, \textit{$d_{out}$}, of all plasmonic devices is kept constant, while the connecting slit, \textit{$w_{slot}$}, has a width of 28~nm. The concept of nanoparticle trapping is also illustrated. (c) Electric field distribution profiles of a single element of the array in the \textit{xy}- and \textit{yz}-planes. The trapping spots appear to be at the walls of the nanoslots with 40~nm width and 80~nm depth.  
   }
\label{fig.1}
\end{figure}

The three-dimensional (3D) nano-aperture platform is illustrated in Fig.~\ref{fig.1}(a).  Nanoring air apertures are engraved on a 50 nm thick gold (Au) metal layer deposited on a quartz substrate. Narrow nanoslot regions are formed by inserting connecting nanogaps between the ring apertures. The plasmonic substrate consists of a 12~(\textit{x}-direction) $\times$ 13~(\textit{y}-direction) array of individual nanoring apertures that are milled using the focussed ion beam (FIB) technique in a 50~nm thick Au film. Scanning electron microscope (SEM) images are used to obtain the dimensions of the nano-apertures, as shown in Fig.~\ref{fig.1}(b). The outer diameter of the nanoring is  \textit{$d_{out}$}=~$285.14~\pm~3.73$ nm, the period of the array, \textit{$\Lambda$} = $361.71~\pm~1.29$ nm, the width of the connecting nanoslots (gap size between the two tips) $w_{slot}=~28.08~\pm~6.04$ nm, and the inner  diameter of the coaxial inner disks, \textit{$d_{in}$}, are 0~nm, $149.59~\pm~2.08$~nm, $173.51~\pm~2.18$~nm, and $195.28~\pm~1.71$~nm. 

The incident light is polarised along the \textit{y}-axis to highly confine the electric field (\textit{E}-field) at the nanoslot area. We use the finite element method (FEM) with the COMSOL Multiphysics software package to simulate the \textit{E}-field distribution and the absorption spectra of the devices. We apply the time-averaged Maxwell's stress tensor~\cite{Stratton} method to calculate the optical gradient force acting on a 20~nm PS particle. The geometric parameters used in numerical simulations are:  $d_{out}=285$ nm, $\Lambda=360$ nm, $w_{slot}=40$ nm, and $d_{in}=0,~150,~175,~200$~nm. We assume that the 20~nm  particles are trapped 5~nm away from the surface of the device. To obtain good resolution for the mesh size at the particle's surface, we choose the nanoslot width to be 20 nm larger than the particle diameter. The  Maxwell's stress tensor is applied on a surrounding shell 2 nm away from the particle's surface. Figure~\ref{fig.1}(c) shows an example of numerical simulations of the \textit{E}-field distribution for the \textit{xy}- and \textit{yz}-planes with a particle positioned at \textit{x} = 0~nm and \textit{z} = 20~nm (the equilibrium position).

\begin{figure}[ht!]
   \centering
   \includegraphics[trim={0cm 2.6cm 0cm 0.5cm},clip, width=12cm, height= 7cm]{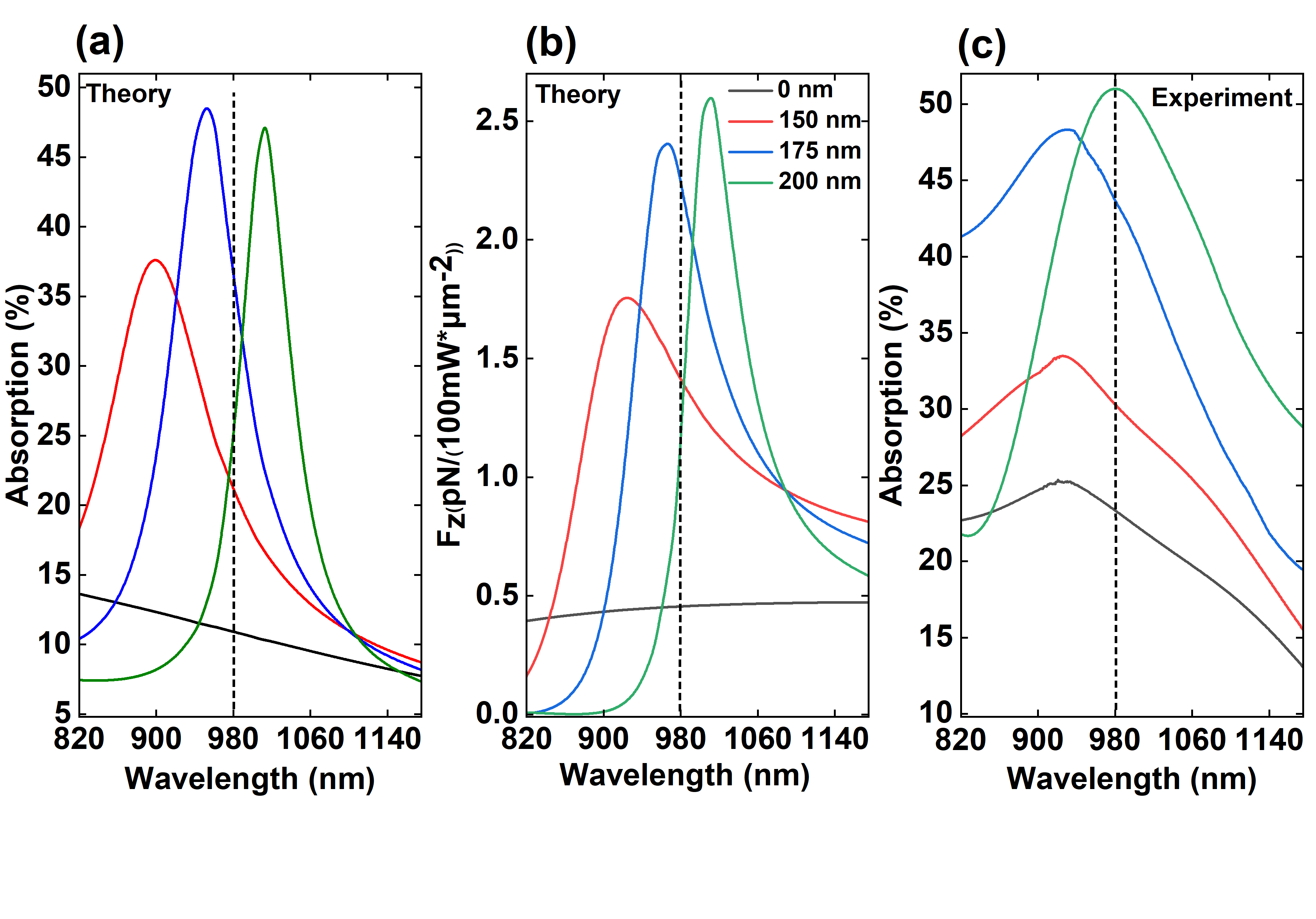}
       \caption{(a) Theoretical absorption spectra for the nano-aperture designs with inner disk diameters 0,~150,~175, and 200~nm. (b) Simulated optical force, \textit{$F_{z}$}, along the light propagation direction, acting on a 20~nm diameter PS particle in water.  The optical force is normalised to 1~mW/$\mu$m$^{2}$ incident light intensity. (c) The absorption spectra of an array of 42 units in deionized water with the polarisation of the microspectrophotometer light source along the \textit{y}-direction. The dotted vertical line indicates the wavelength of the laser in trapping experiments. 
       }
\label{fig:2}
\end{figure}

Absorption peak spectra are used as indicators of the resonance positions for the various structures used in the study. Figure~\ref{fig:2}(a) shows the theoretical absorption spectra for all the structures with varying inner disk diameters, \textit{$d_{in}$} = 0 - 200 nm, in water solution. Note that, with increased inner disk size, the absorption peak is red-shifted, which allows us to tune the required excitation laser towards the experimental wavelength of 980~nm. Of the four devices, the 175~nm inner disk design shows the highest theoretical absorption peak at 980 nm (vertical dotted line in Fig.~\ref{fig:2}(a)).  In Figure~\ref{fig:2}(b), we plot the theoretical optical trapping force along the \textit{z}-axis (light propagation direction), acting on a 20~nm PS nanoparticle at ($x,y,z)~=~(0,0,-4)$~nm position,  
where the localised field intensity is highest, as a function of the excitation laser wavelength. For  excitation  at 980~nm, the 175~nm design exerts the strongest theoretical optical force, of about 2.24~pN/(100 mW$\cdot$ $\mu$m$^{-2}$), on the nanoparticle. We see that the theoretical absorption spectra and $z$-component of the optical trapping force show similar curve trends. However, the theoretical trapping force peak positions show a slight red-shift towards the 980 nm excitation wavelength compared to the theoretical absorption resonance peaks. It is worth noting that the optical force is relatively constant for the 0~nm design (no inner disk) within this wavelength range.

To perform the experimental measurement of absorption, a microspectrophotometer (MCRAIC 20/30 PV) is used to measure the reflection, \textit{R}, and transmission, \textit{T}, spectra of the plasmonic devices in deionized water. The absorption, \textit{A}, is determined from the following : ${A(\%)} = 100({\%})-T(\%)-R(\%)$ and indicates the resonance peak position for the nanoring array structures, see Fig.~\ref{fig:2}(c). We see that the wavelength is blue-shifted by approximately 20~nm compared to the theoretical curves (Fig.~\ref{fig:2}(a)). We will discuss this discrepancy in the following sections. Moreover, absorption spectra artifacts were observed for inner disk diameters of 0~nm, 150~nm, and 175~nm. This is a systematic error caused by the detector changes (from the visible to the near-infrared regime) and leads to spurious peaks on the absorption spectrum, (Fig.~\ref{fig:2}(c)). However, the experimental trapping laser beam at 980~nm is at a value far from these  peaks. This allows us to extract information on the enhancement of the absorption spectrum for each plasmonic configuration.

\subsection{Optical potential well and trap stiffness}

The optical trapping force, its corresponding potential, and trap stiffness for a 20~nm PS particle  in the proximity of the 3D  plasmonic nanoslot area are numerically calculated using the COMSOL Multiphysics software package. We assume that the particle is in the water domain of the \textit{xy}-plane and sweep the particle along the \textit{z}-axis (see Fig.~\ref{fig.1}(b)). We obtain the potentials by integrating the corresponding forces along the \textit{z}-axis. The incident laser intensity is 100~mW/$\mu$m$^{2}$ at 980~nm.

\begin{figure}[htp]
\centering
\includegraphics[trim={2cm 0.6cm 3cm 1.5cm},clip, width=6cm, height=5.2cm]{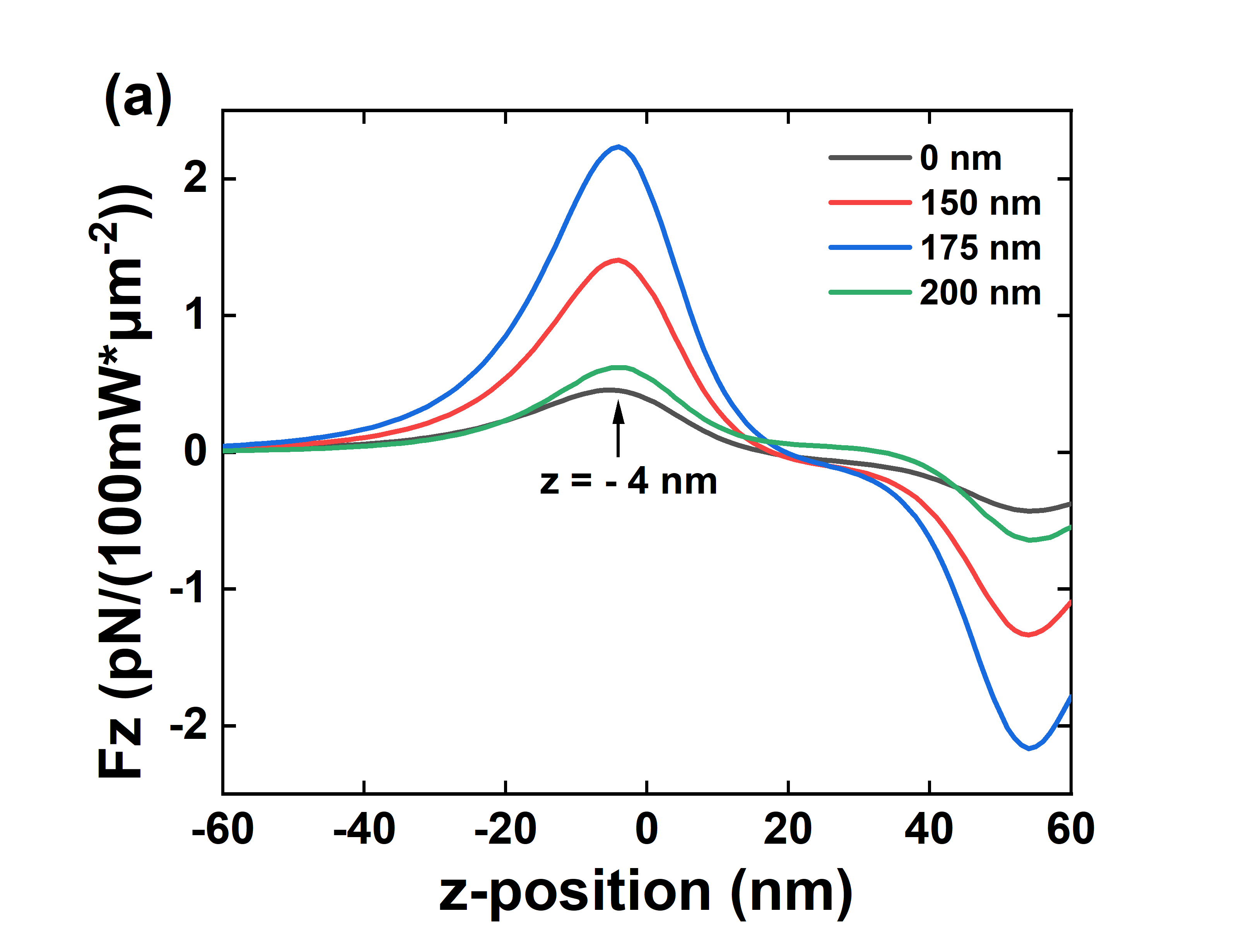}
\includegraphics[trim={1.3cm 0.7cm 3cm 1.2cm},clip,width=6cm, height=5.2cm]{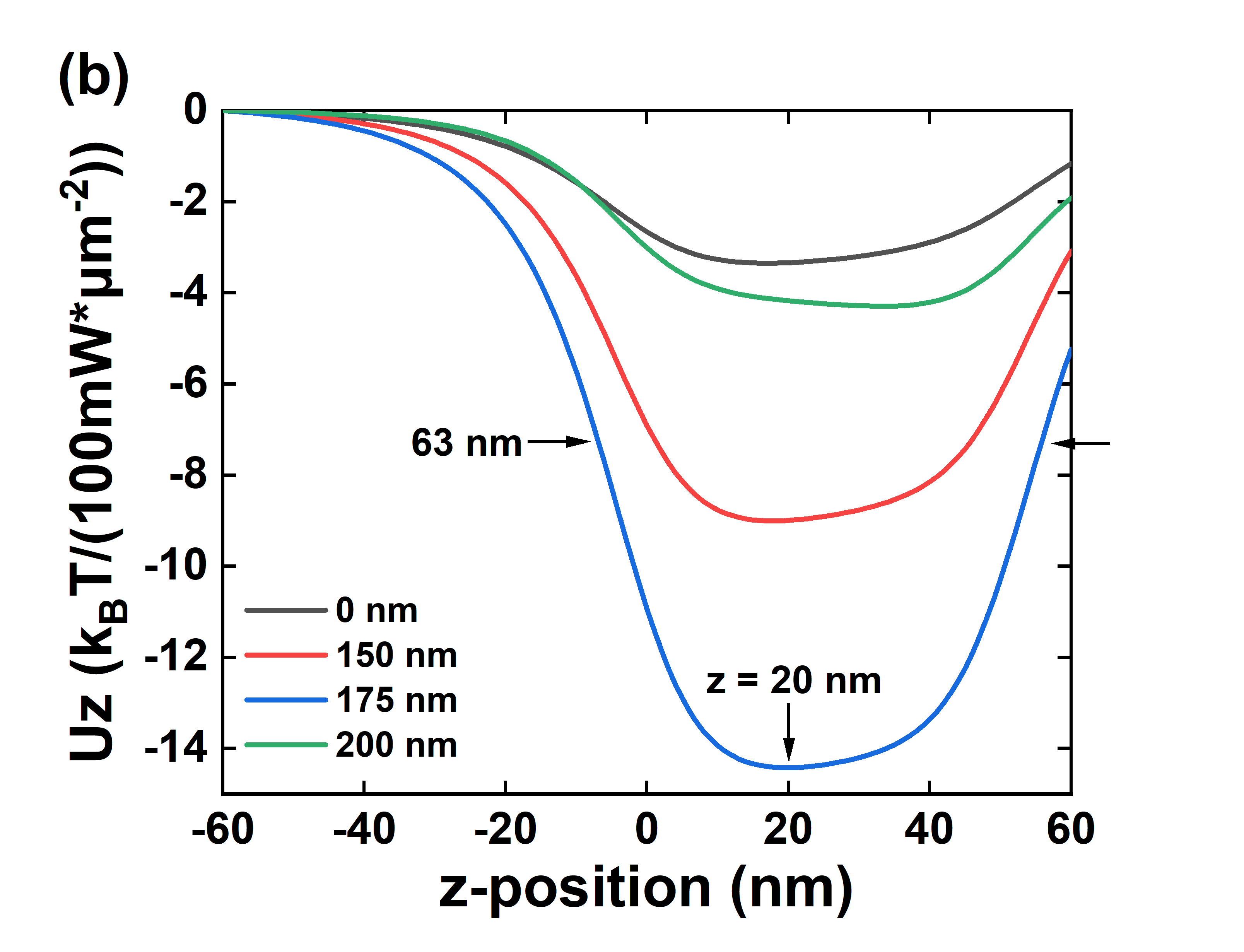}
\caption{\label{force_potential_all}(a)
Optical forces along the \textit{z}-axis for 980~nm illumination. 
The maximum optical force is at \textit{z} = -4~nm for all  designs. (b)
Corresponding potentials of the optical forces in (a). The depths of the  wells have a similar behaviour as the optical forces, with the 175 nm inner disk diameter design exhibiting the deepest and broadest well, with a full-width-at-half-maximum of 63~nm. The equilibrium position is around \textit{z} = 20~nm.}
\end{figure}

In our model, the plane at \textit{z} = 0~nm is the boundary between the Au film and the water domain. The thickness of the Au film is 50~nm, thus for positions in the range of 0~-~50~nm along $z$ the particle is  within the water-filled nanoslot region of the Au film. The optical force has a positive sign when the particle is attracted towards the SiO$_2$ domain and a negative sign indicates a repulsive force in the opposite direction. Moreover, the equilibrium position for trapped particles is located at \textit{z} = 20~nm, i.e., inside the  nanoslot area, for all  designs. Figures~\ref{force_potential_all}(a) and (b) show the calculated optical forces and their corresponding potentials  along the \textit{z}-axis for four different nanodisk structures.

We also calculate the trapping potentials  along the \textit{x}- and \textit{y}-axis for the 175~nm inner disk diameter and at the \textit{z} = -4~ nm plane, where the strongest optical force is observed (data not shown). We obtain a full-width-at-half-maximum (FWHM) of the trapping  wells in the \textit{x}-direction to be 44~nm  and  14.4~nm in the \textit{y}-direction, which is almost four times narrower due to stronger
trapping. This arises since the \textit{E}-field is highly confined (i.e., stronger gradient) at the nanoslot along the \textit{y}-direction.  

Finally, the trap stiffness, \textbf{$\kappa$}, is calculated for the 175~nm inner disk design, along all three axes. We use the harmonic oscillator equation, $F=\kappa \Delta r$, where \textit{$F$} is the maximum optical force acting on the nanoparticle for each axis and \textit{$\Delta r$} is the FWHM/2. We also normalise all values to 1~mW/$\mu$m$^{2}$ incident light intensity. We determine that $\kappa_x = 0.55$~fN/nm, $\kappa_y=~3.72$~fN/nm, and $\kappa_z = ~0.71$~fN/nm. The total trap stiffness is  $\kappa_{tot}=\sqrt{\kappa_x^2 + \kappa_y^2 + \kappa_z^2}=3.82$~fN/nm. 

The structure with  the 175~nm inner disk diameter has the highest theoretical absorbance efficiency, provides the maximum theoretical optical force on a 20 nm particle, and creates the deepest trapping potential of -14.42$k_{B}$T/(100~mW/$\mu$m$^{2}$). As mentioned previously, there is a 20 nm blue-shift of the experimental absorption peak positions compared to theoretical calculations. We observe that the fabricated 200 nm design nanodisk shows the strongest experimental absorbance at 980 nm (vertical dashed line in Fig.~2(c)). We assume that this discrepancy arises from imperfections during the fabrication process leading to rounded edges in the structure's features, in contrast to sharp edges in simulations. Based on theoretical interpretations, we  conclude here that the 175 nm nanodisk structure should provide the strongest experimental optical trapping force on a 20 nm PS particle in water solution for the excitation wavelength of 980nm.

\subsection{Trap performance evaluation}
The experimental approach is described in our previous work \cite{R4}. In brief, for nanoparticle trapping experiments, a tunable, continuous-wave (cw) Ti:Sapphire laser (MBR110 Coherent) is used at 980 nm. The laser spot diameter at the sample plane is around 2 $\mu$m. The number of nanorings that can be excited on the plasmonic substrates is 25, based on the size of the incident trapping laser spot. A high numerical aperture (N.A.= 1.3) oil immersion objective lens (OLYMPUS UPlanFL N 100$\times$) is used to focus the trapping beam onto the plasmonic substrate. We control the incident power of the trapping beam, which is limited to a maximum of 7.7 mW, at the sample plane, while the polarisation of the trapping laser is along the \textit{y}-direction (see Fig.~\ref{fig.1}). A trapping event is detected by collecting the transmitted laser light through a 50$\times$ objective lens (Nikon CF Plan) and sending it to an avalanche photodiode (APD430A/M, Thorlabs). The APD signal is recorded using a data acquisition board at a frequency of 100 kHz with a LabVIEW program. When a 20 nm particle is trapped in one connecting nanoslot, a sharp increase in transmission is observed. The transmission level returns to its initial value when there is no particle  at the trapping point. The plasmonic substrate is attached to a cover glass with adhesive microscope spacers, forming a microwell. The microwell contains 8 $\mu$L of PS nanoparticles with a mean diameter of 20 nm (Thermo Fisher Scientific, F8786) in heavy water  with a concentration of 14~$\times$~10$^{11}$ particles/ml. A small amount of surfactant (Detergent Tween 20 with 0.1\% volume concentration) is used to minimize particle aggregation. The microwell is mounted and fixed on top of a piezoelectric translation stage.

\begin{figure}[htp]
\centering
\includegraphics[width=0.49\textwidth]{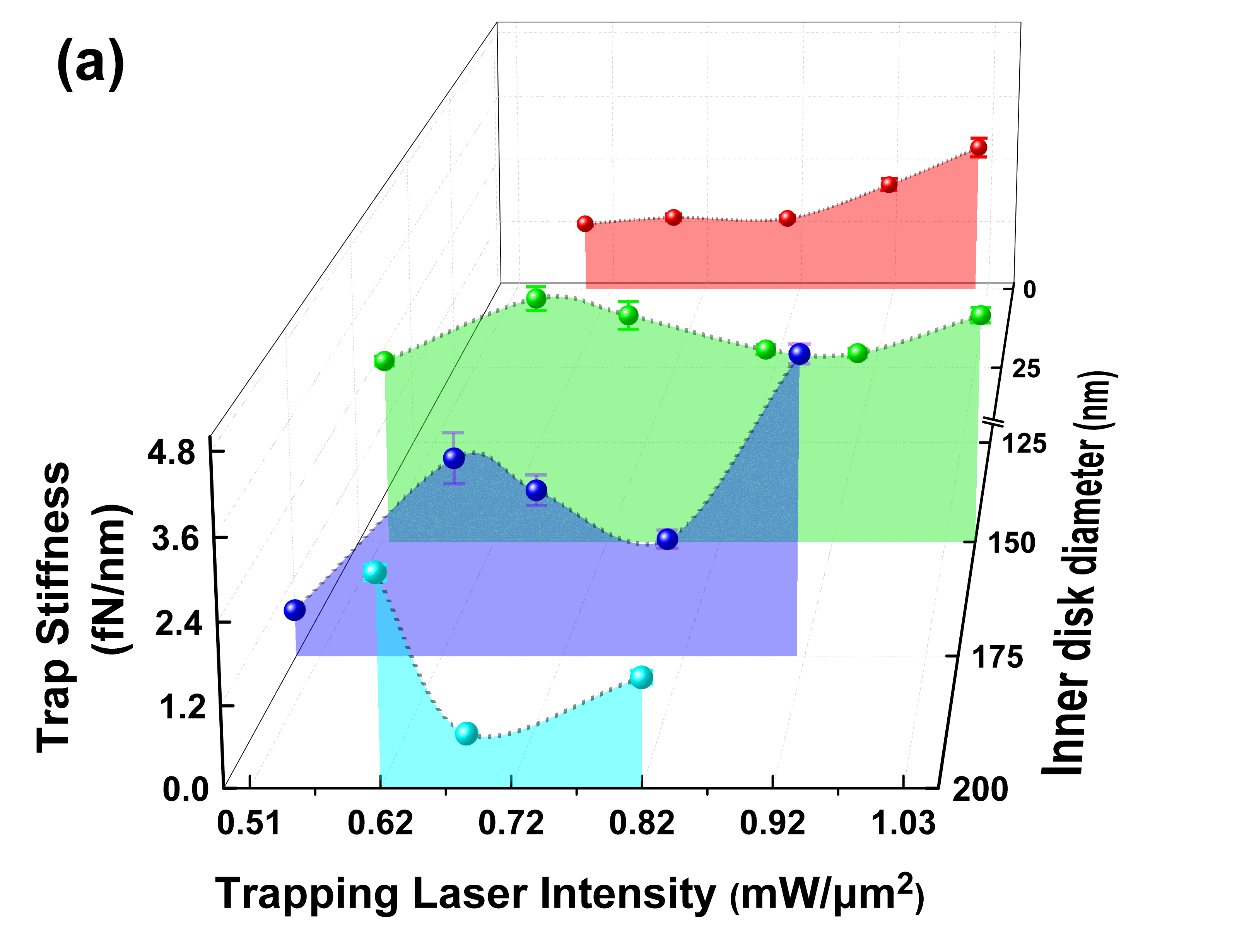}
\includegraphics[ width=0.49\textwidth]{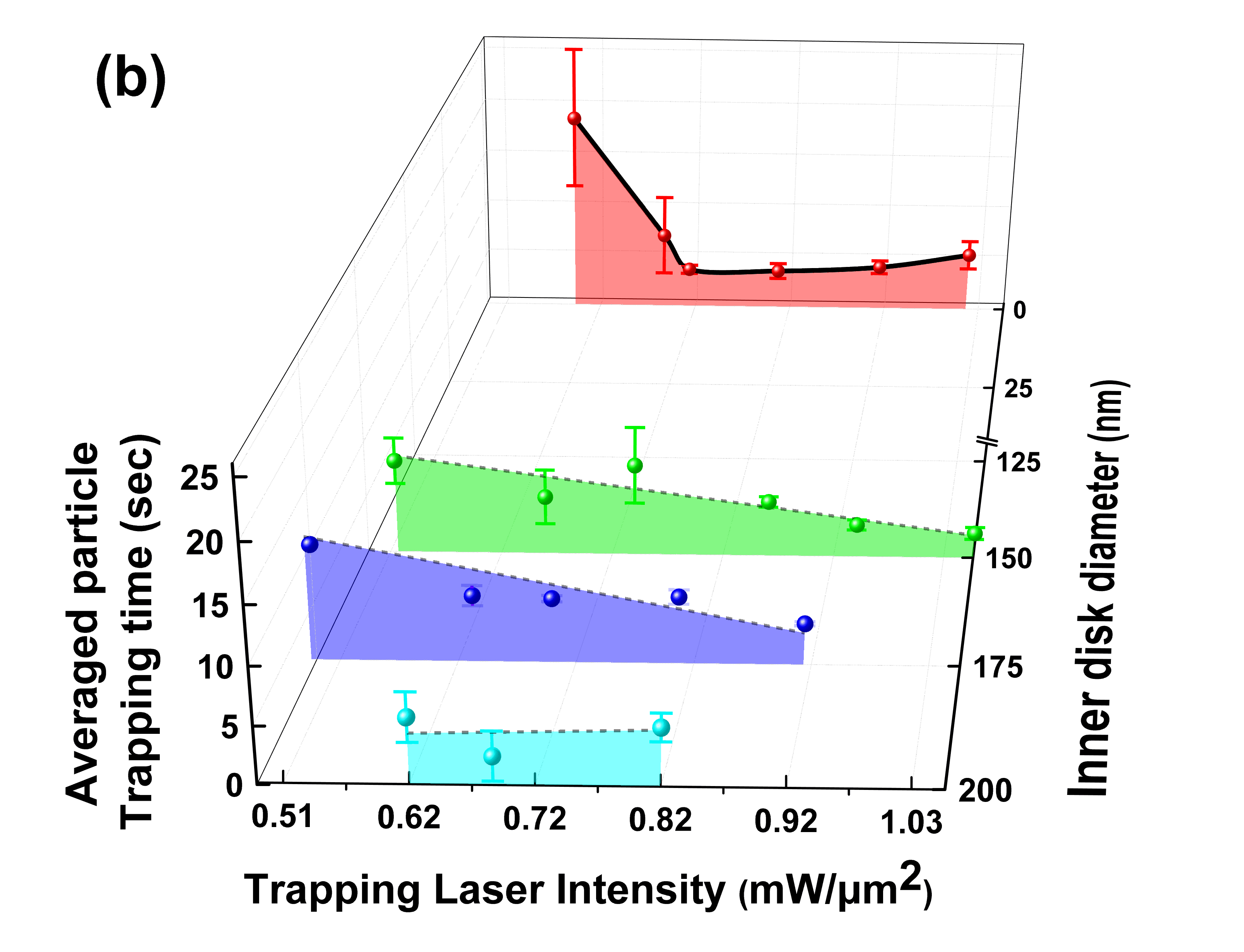}
\includegraphics[width=0.49\textwidth] {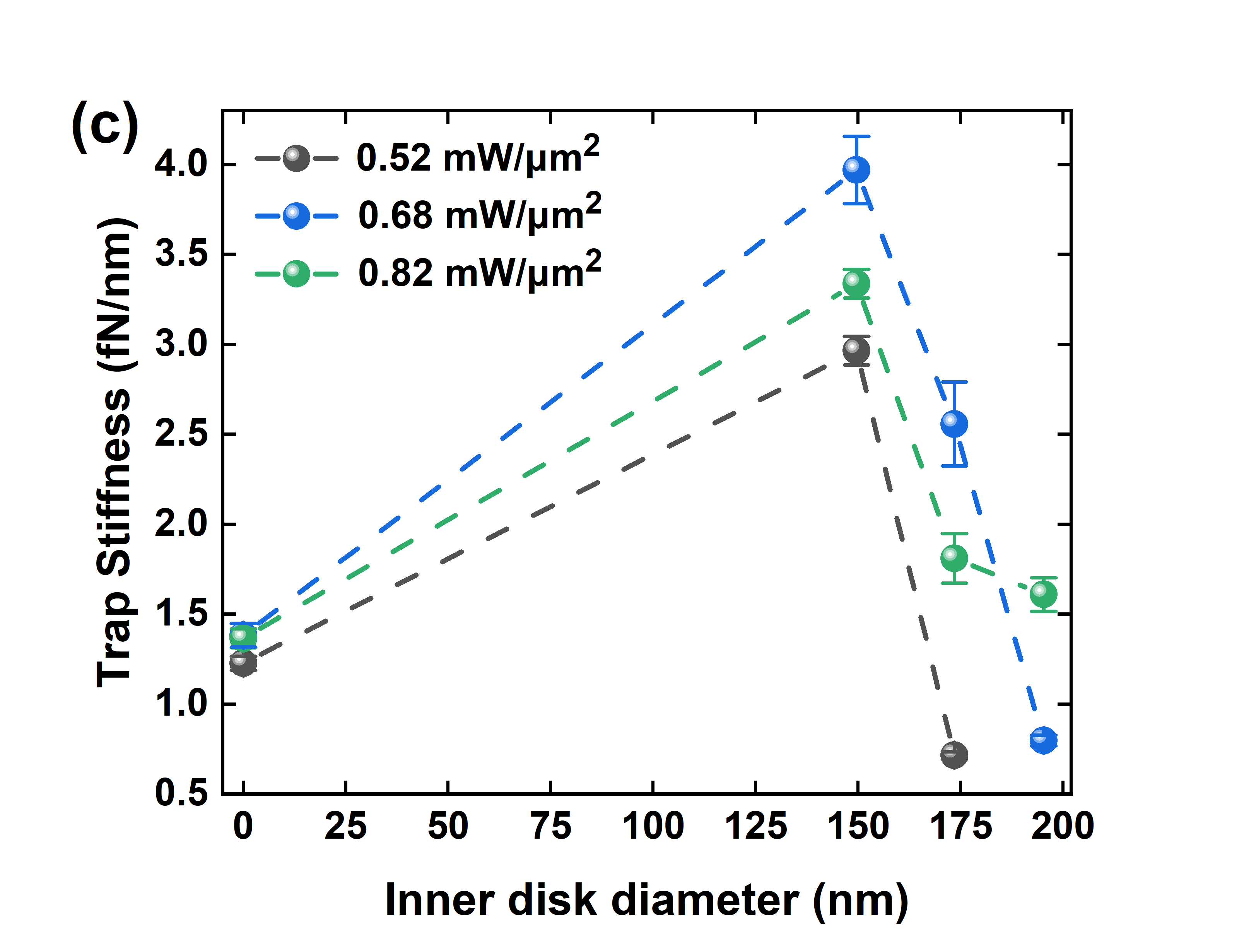}
\includegraphics[width=0.49\textwidth]{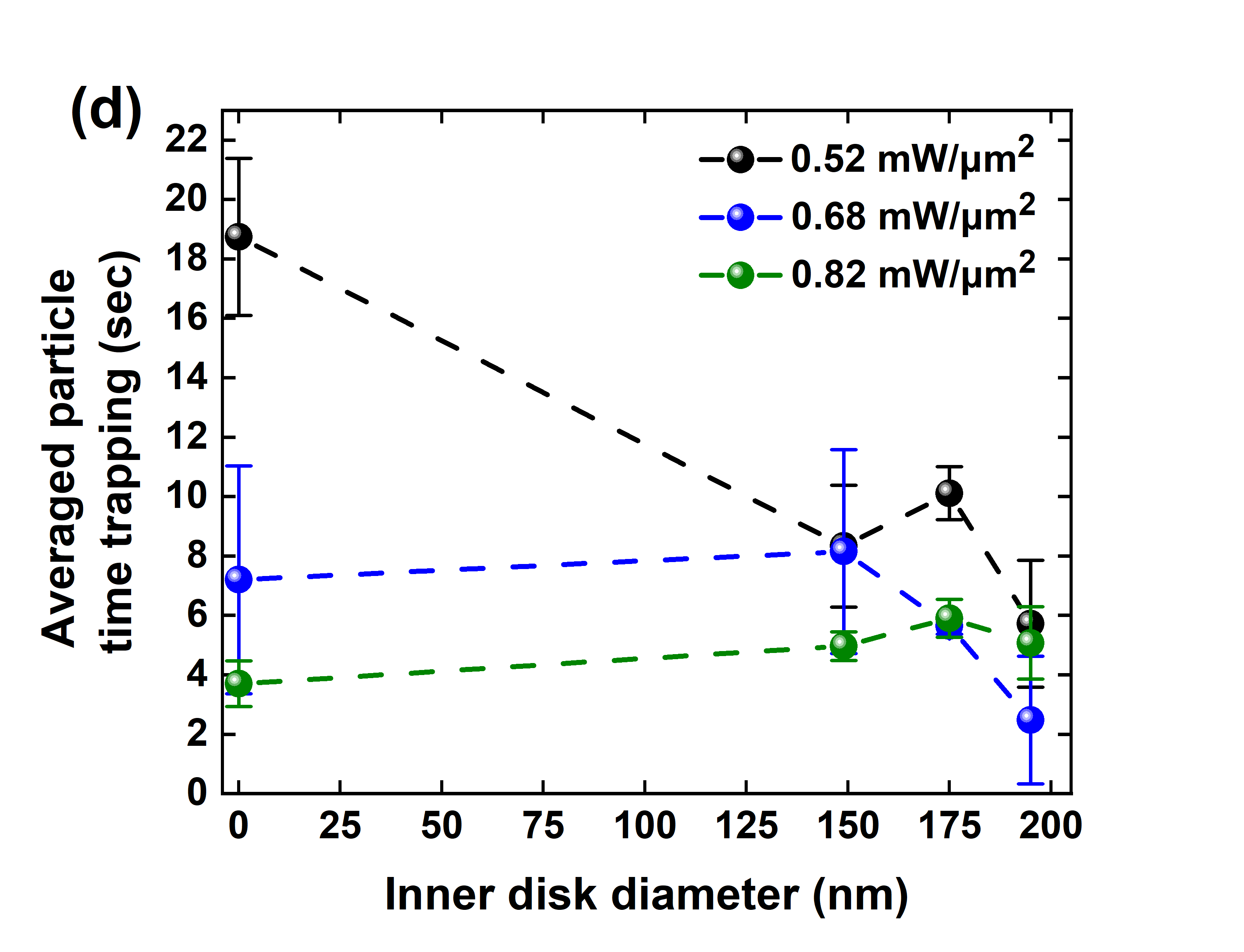}
\setlength\abovecaptionskip{10pt}
\caption {(a) Trap  stiffness  for  a  20~nm PS particle  as  a  function  of  incident  laser intensity and inner disk diameter. The dotted lines serve as visual aids. (b) Average particle trapping time as a function of trapping laser intensity and inner disk diameter. The black solid line for 0~nm inner disk diameter serves as a visual aid while the dashed lines are  linear fits to  the experimental results. Each color represents a different disk diameter. (c) Trap stiffness and (d) average time to trap a particle as a function of inner disk diameter for 0.52~mW/$\mu$m$^{2}$, 0.68~mW/$\mu$m$^{2}$, and 0.82~mW/$\mu$m$^{2}$ incident laser intensities. The points plotted are averaged values over three trapping events for different laser intensities (different colors), along with their standard deviations. 
}
\label{fig:4}
\end{figure}

Figure \ref{fig:4}(a) shows the measured trap stiffnesses, \textit{$\mathrm{\kappa}_{exp}$}, for four different inner disk diameters, \textit{$d_{in}$}, as a function of the incident laser intensity. We use the transient time method as discussed elsewhere \cite{Kotnala,Han, R4}. We also consider Fax\'{e}n's correction to introduce a factored drag force that arises due to the surface roughness of the nanoring structure wall, as previously reported for nanoparticle trapping \cite{Kotnala, Han, R4}. The trapping laser intensity ranges from 0.48~mW/$\mu$m$^{2}$ to 1.03 mW/$\mu$m$^{2}$. The experimental trap stiffness is the average over three multiple runs for each laser intensity. It should be noted that the range of the laser intensity varies for each configuration due to difficulties in achieving single particle trapping.  

As expected, we observe an increase to the  measured trap stiffnesses for larger inner disk structures compared to the no-disk configuration. When there is no disk (red curve in Fig.~\ref{fig:4}(a)), the trap stiffness is relatively constant for trapping laser intensities below 0.75~mW/$\mu$m$^{2}$ and increases linearly for higher trapping laser intensities. This behaviour changes for the nanoring array with \textit{$d_{in}$} = 149 nm (green curve in Fig. \ref{fig:4}(a)), where the trap stiffness does not vary significantly as the incident laser intensity increases to 1.03 mW/$\mu$m$^{2}$. Surprisingly, as the inner disk diameter increases, larger fluctuations of the trap stiffness are evident and the expected linear behaviour is not observed. A maximum trap stiffness is achieved for the 149 nm inner disk configuration (purple curve in Fig. \ref{fig:4}(a)). We also note that the 195 nm inner disk diameter device (blue curve in Fig~\ref{fig:4}(a)), which is expected to produce the highest optical trapping force, does not provide the maximum trap stiffness for both low, 0.52 mW/$\mu$m$^{2}$, and high, 0.82 mW/$\mu$m$^{2}$, laser intensities, (Fig.~\ref{fig:4}(c)). Instead, the trap stiffness for larger inner disk diameters of 173 nm and 195 nm decreases compared to the 149 nm inner disk structure. Assuming that, as the inner disk diameter increases, thermal energy dissipation via the Joule effect increases upon resonant excitation, an increase of the spatial temperature distribution would result. This is due to collective heating of several hotspots in the illumination area of the nano-aperture array. This photo-induced plasmonic heating produces a temperature gradient across the device, resulting in the creation of a strong thermal fluidic flow, which could influence the trap stiffness values.

The maximum  $\kappa_{exp}=3.50\pm0.17$ fN/nm is obtained for the nanoring array with 149~nm inner disk diameter. The stiffness enhancement factor is approximately 2.9, 1.8, and 0.6 times for the nanoring arrays of \textit{$d_{in}$} = 149 nm, \textit{$d_{in}$} = 173 nm, and \textit{$d_{in}$} = 195 nm, respectively, when compared to the standard, \textit{$d_{in}$} = 0 nm, nanohole array configuration for an incident laser intensity of 1.0~mW/$\mu$m$^{2}$ at 980 nm. The high trap stiffness for the  149~nm inner disk diameter indicates the ability of this device to perform stable trapping and  to hold  nanoparticles at specific positions for long periods of time, thereby providing opportunities for effective low-power nanomanufacturing of nanoscale objects.

Figure~\ref{fig:4}(b) shows the average time taken to trap a single 20 nm particle as a function of the trapping laser intensity for different inner disk diameters. The trapping time is defined as the time from when the trapping laser is turned on (\textit{t} = 0) to the transmission signal's first observed discrete step, indicating the first particle trapping event. The experimental trapping time is taken from the average values observed over three multiple runs. The error bars represent the standard deviation of the average trapping time measurements. We also assume that, as the trapping laser intensity increases, the spatial temperature distribution increases, producing buoyancy-driven natural convection due to the fluid’s density gradient. This leads to an increase in the probability of trapping a nanoparticle in the flow, delivering it to the illuminated area, thus reducing the average particle trapping time. We also note that, for low laser intensities, the array of nano-apertures (0~nm inner disk) provides the longest particle diffusion times (Fig.~\ref{fig:4}(d)). Specifically, for 0.55~mW/$\mu$m$^{2}$ laser intensity, the trapping time for the 0~nm and 149~nm inner disk designs is 18.7~sec and 8.3~sec, respectively. A linear behaviour of the average particle trapping time as a function of laser intensity is observed for configurations with 149~nm and 173~nm inner disk diameters with slopes of 1.72~$\pm$~0.35~s/(mW/$\mu$m$^{2}$) and 2.53~$\pm$~0.65~s/(mW/$\mu$m$^{2}$), respectively. Figure~\ref{fig:4}(d) shows that the average particle trapping time approaches a minimum value at around 4 seconds for all nanodisk designs when higher laser intensities up to 0.82~mW/$\mu$m$^{2}$ are used.

\section{Conclusion}

 Heating in plasmonic tweezers has previously been treated as an obstacle to stable trapping of particles. In 2010, Ploschner et al. performed a computational study of trapping with a plasmonic nano-antenna and suggested that the particle positioning may not be due to optical forces alone~\cite{R33}. The discrepancies we observe in the values of trap stiffness determined from theory based on  optical force calculations and  experimental observations indicate that thermophoresis may play a significant role in the trapping process for larger inner disks thereby creating an additional force on the particle due to Stokes' drag. To further understand the origin of this mechanism, a 3D theoretical heating model of plasmonic arrays is required to explore in depth the contribution of thermal convection flows to the trapping performance. A detailed study of particles (e.g., biomolecules, dielectric, or metallic nanoparticles with varying particle concentrations and sizes) will provide insights into the thermally assisted trapping process. Besides these parameters, other benchmarks,  such as interactions between particles and the surface, should be addressed for a complete understanding of  the experiments. This is beyond the scope of the current work.
 
 In 2014, Roxworthy et al. showed that an array of plasmonic nano-antennas coupled to an optically absorptive indium-tin-oxide (ITO) substrate can generate micrometre per second fluid convection~\cite{R32}. In their system, heating creates a fluid motion capable of rapid particle transport. Here, we replace the absorbing ITO thin film substrate with an array of plasmonic  nano-apertures, engraved directly onto a 50~nm metallic Au film. The Au metal film itself facilitates the thermal energy distribution of the Au substrate and the surrounding fluid;  hence, heat transfer can occur over large effective areas of the illuminated aperture array. Collective heating of many hotspots achieves better heating efficiency at relatively low laser power excitation compared to a single nano-element trapping configuration. This, in turn, produces buoyancy-driven natural convection currents and can rapidly transport the nanoparticles towards hotspots where they will be trapped in a short time interval.   Our approach provides an alternative avenue towards rapid particle delivery and strong single nanoparticle trapping, both of which could play important roles in molecular analysis.
 
\subsection*{Acknowledgement}The authors would like to thank M. Ozer and S. P. Mekhail for technical assistance, M. Sergides for initial contributions to the experimental work, and  OIST editing section for reviewing the manuscript.

\subsection*{Funding}
This work was partly funded by the Okinawa Institute of Science and Technology Graduate University, Onna-San, Okinawa, Japan. DGK acknowledges support from JSPS Grant-in-Aid for Scientific Research (C) Grant Number GD1675001.

\bibliography{manuscript}
\end{document}